\documentstyle[12pt,aaspp4,flushrt]{article}
\doublespace
\begin{document}

\title{The Thermal Response of a Pulsar Glitch :\\
		The Non-spherical Symmetric Case}

\author{K.S. Cheng$^1$, Y.Li$^{1,2}$ and W.M.Suen$^3$}
\affil{
$^1$Department of Physics, University of Hong Kong, Hong Kong, China\\
$^2$High Energy Astrophysics Laboratory, IHEP, Beijing, China\\
$^3$Department of Physics, Washington University, U.S.A.
}
\begin{abstract}

We study the thermal evolution of a pulsar after a glitch in which the
energy is released from a relative compact region.  A set of
relativistic thermal transport and energy
balance equations is used to study the thermal evolution, without making
the assumption of spherical symmetry.  We use an
exact cooling model to solve this set of differential equtions.  Our
results differ significantly from those obtained under the assumption
of spherical symmetry.  Even for
young pulsars with a hot core like the Vela pulsar, we find that a
detectable hot spot can be observed after a glitch.  The results suggest that 
the intensity variation and the relative phases of hard X-ray emissions in 
different epoches can provide important information on
the equation of state.
\end{abstract}

\keywords{
dense matter - stars: evolution - star: interiors - stars: neutron - stars: X-rays}

\section{Introduction}
The studies of the thermal evolution of pulsars are believed to provide
vital information on the internal properties of the neutron stars 
(for a review see e.g., Tsuruta 1992).  Theoretical cooling curves are 
often compared with the surface temperatures of pulsars with 
different ages.  However, different pulsars clearly have different 
parameters, e.g. the rotation period ($P$) 
and the surface magnetic field ($B$), 
which may affect the cooling processes.  Several proposed 
internal heating mechanisms, e.g. the fractional heating between 
the crustal superfluid and the crust (Alpar et al 1984a, Shibazaki 
and Lamb 1988), the crust cracking (Cheng et al 1992), the chemical 
heating (Reisennegger 1995) etc., all depend on the pulsar parameters, 
i.e. $P$ and $B$.  Furthermore, different pulsars 
have different internal properties, e.g. mass, 
equation of state, 
impurity content etc., which strongly affect the cooling curve.  
Hence ideally speaking, one would like 
to study the thermal evolution process 
of an individual pulsar.  But realistically, 
the normal cooling process is an 
extremly slow process and cannot be followed observationally 
for an individual star. 

The thermal evolution of a pulsar after a glitch is particularly
interesting in that it fills in this gap.  A large amount of energy
could be suddenly released during the period jump (glitch).  The
energy will eventually be transported to the surface of the star, and
released in the form of a transient thermal X-ray emission.  Several
authors (Van Riper et al 1991, Chong and Cheng 1994, Hirano et al
1997) have calculated the thermal evolutions of pulsars after glitches,
but they all assumed that the energy is generated in a spherical shell
inside the star.  They found that the glitches cannot produce very
significant observed results for young pulsars like the Vela pulsar
(Van Riper et al 1991).  However, the energy released in
the glitch is likely to be deposited in a compact region in the inner crust,
by either superfluid unpinning (e.g. Cheng et al 1988, Alpar and Pines
1995) or crust cracking (e.g. Ruderman 1991, Cheng et al 1992).  The
energy transport is clearly not spherical symmetric (although the
background geometry is to a good approximation spherical).  In this
paper, we study the general case of the energy transport inside
neutron stars after a glitch without the spherical symmetry
assumption, and arrive at a conclusion substantially different from
previous results.  In section 2, we
derive the general relativistic non-spherical symmetric transport and
energy balance equations. The necessary physics inputs and solution
algarithms for solving this set of relativistic differential equations
are decribed in section 3. Numerical results 
are presented in section 4, with a brief discussion in section 5.

\section{General Relativistic Non-Spherical Symmetric 
Thermal Transport and
Energy Balance Equations}

The Newtonian thermal transport and 
energy balance equations are given respectively by
\begin{equation}
\label{eq01}
\nabla\cdot {\bf F}=-C_{V}{dT\over dt}=-nT{ds\over dt}-Q_{\nu}
\end{equation}
and
\begin{equation}
\label{eq02}
\nabla T=-{{\bf F}\over K}
\end{equation}
where $\bf F$ is the energy flux, 
$C_{V}$ is the specific heat capacity, 
$T$ is the temperature, 
$n$ is the particle number density, 
$s$ is the specific entropy per particle, 
$Q_{\nu}$ is the neutrino emissivity per unit volume and 
$K$ is the thermal conductivity.  

To generalize the above equations to relativistic situation, we 
make the following assumptions. (1) The neutron star is rotating slowly 
enough that the metric tensor describing the 
background spacetime can be written as (Tolman 1934)
\begin{equation}
\label{eq03}
ds^2=- e^{2\Phi(r)} dt^2 + e^{2\Lambda(r)}dr^2 + r^2(d{\theta}^2+
sin^2{\theta}d{\varphi}^2)\;\;,
\end{equation}
where $e^{\Lambda(r)}=(1-2Gm/rc^2)^{-1/2}$.  (2) The 
diffusion limit is appropriate for the energy transport.  (3) There 
are no other entropy-generating mechanism besides diffusion, and 
second order flux terms in the transport are neligible. (4) Fluid motion 
inside the star due to the thermal effects is negligible.  
These simplifying assumptions are clearly justified for the problem at 
hand.  The energy transport is then governed by the following 
equations.  The energy-momentum tensor can be written as
\begin{equation}
\label{eq04}
T^{\mu\nu}=(\rho+P)U^{\mu}U^{\nu}+\rho g^{\mu\nu}+
q^{\mu}U^{\nu}+U^{\mu}q^{\nu}
\end{equation}
where
$U^{\mu}$ is the 4-velocity of fluid flow, 
$\rho$ is total energy density measured in 
the rest frame of fluid, and 
$P$ is the presure in rest frame of fluid.   
The heat flow is given by 
$q^{\mu}=-K(g^{\mu\nu}+U^{\mu}U^{\nu})(T,_{\nu}+Ta_{\nu})$
where 
$a_{\nu}=U^{\mu}U_{\nu;\mu}$ 
is the 4-acceleration, and $K=K(T)$ is the effective conductivity,
e.g. $K\propto T^3(t,r,\theta,\varphi)$ for photon
diffusion.  Since we assume that there is no fluid motion
the 4-velocity of the fluid is 
$U^{\alpha}$=(e$^{-\Phi}$,0,0,0) and 
$a^{\alpha}$=(0,e$^{-2\Lambda}\partial_r\Phi$,0,0). 
The 4-heat flow is given by 
$q^{\alpha}=(0,-Ke^{-2\Lambda}(\partial_rT+T\partial_r\Phi),
-{K\over{r^2}}\partial_{\theta}T,
-{K\over{r^2sin^2\theta}}\partial_{\varphi}T)$.  
The energy-momentum tensor components related to 
the energy transport equation are following:
$T^{tt}=\rho U^tU^t$=$\rho e^{-2\Phi}$,
$T^{tr}=f^r=q^rU^t$=$-Ke^{-\Phi-2\Lambda}(\partial_rT+T\partial_r\Phi)$,
$T^{t\theta}=f^{\theta}=q^{\theta}U^t$=$-{Ke^{-\Phi}\over{r^2}}
\partial_{\theta}T$
and 
$T^{t\varphi}=f^{\varphi}=q^{\varphi}U^t$=$-{Ke^{-\Phi}\over{r^2sin^2\theta}}
\partial_{\varphi}T$.
To express the thermal transport equation 
in terms of local observables, we note that
$T^{\hat{\alpha}\hat{\beta}}=\frac{\partial\zeta^{\hat{\alpha}}}
{\partial x^{\alpha}}\frac{\partial\zeta^{\hat{\beta}}}{\partial
x^{\beta}}
T^{\alpha\beta}$ 
where $d\zeta^{\hat{\alpha}}$
=($e^{\Phi}dt, e^{\Lambda}dr, rd\theta, rsin{\theta}d\varphi$) 
are the unit one forms, 
and $dx^{\alpha}$=($dt, dr, d\theta, d\varphi$).  
The locally measured energy flux in spherical coordinate is
\begin{equation}
\label{eq10}
{\bf F}=(T^{\hat{t}\hat{r}},
T^{\hat{t}\hat{\theta}},T^{\hat{t}\hat{\varphi}})
\end{equation}
where
\begin{equation}
\label{eq11}
T^{\hat{t}\hat{r}}=e^{(\Lambda+\Phi)}f^r=
-Ke^{-\Phi-\Lambda}\partial_r(e^{\Phi}T),
\end{equation}
\begin{equation}
\label{eq12}
T^{\hat{t}\hat{\theta}}=re^{\Phi}f^{\theta}=-{K\over{r}}\partial_{\theta}T
\end{equation}
and 
\begin{equation}
\label{eq13}
T^{\hat{t}\hat{\varphi}}=e^{\Phi}rsin\theta f^{\varphi}
=-{K\over{rsin\theta}}\partial_{\varphi}T.
\end{equation}
Therefore, the relativistic thermal transport equation is given by
\begin{equation}
\label{eq14}
(e^{\Phi}T)_{;i} =
-{e^{\Phi}{F_{i}}\over {K}}
\end{equation}
where $;i$ denotes spatial covariant derivative on the constant time
slice of the metric given by Eq.(\ref{eq03}), 
with $i= {r, \theta , \varphi}$.  
We can see that this equation (i) in the Newtonian limit reduces to
to Eq.(\ref{eq02}) with $\Phi$ 
and $\Lambda$ go to zero, and (ii) in the spherical symmetric case reduces
to the equations of, e.g., Straumann (1984), with  
$\partial\over{\partial \theta}$ 
= $\partial\over{\partial \varphi}$ = 0. 

The energy balance equation can be derived by the conservation of
the energy-momentum tensor, namely
$0 = T^{t\mu}_{~~;\mu} = T^{tt}_{~~;t}+{2\over r}f^r+
{cos\theta\over{sin\theta}}f^{\theta}+
\partial_rf^r+f^r\partial_r\Lambda
+3f^r\partial_r\Phi+\partial_{\theta}f^{\theta}+
\partial_{\varphi}f^{\varphi}$ 
where 
$T^{tt}_{~~;t}={\partial\rho\over{\partial t}}e^{-2\Phi}
={\partial\rho\over{\partial\tau}}e^{-\Phi}$. Here,
$\partial\rho\over{\partial\tau}$ is the rate of change of energy
density measured in proper frame.  $\partial\rho\over{\partial t}$
depends on the processes under consideration, e.g., if only heat
conduction is considered, one has ${\partial\rho\over {\partial t}}=C_V{\partial
T\over{\partial t}}$, with $C_V$ being the heat capacity measured in the
proper frame.  We include neutrino emission, which leads to
${\partial\rho\over {\partial t}}=C_V{\partial T\over{\partial
t}}+e^{\Phi}Q_{\nu} =
{e^{-\Phi-\Lambda}\over{r^2}}
\partial_r(r^2f^re^{\Lambda+3\Phi})+
{e^{2\Phi}\over{sin\theta}}
\partial_{\theta}(sin\theta f^{\theta})+
e^{2\Phi}\partial_{\varphi}f^{\varphi}$.  
Using the relations between $f^{\alpha}$ and 
$F^{\alpha}$ in Eqs (\ref{eq11}) to (\ref{eq13}), 
we obtain
\begin{equation}
\label{eq15}
-(C_V{dT\over dt}+e^{\Phi}Q_{\nu})=
{e^{-(\Phi+\Lambda)}\over r^2}{\partial\over{\partial
r}}(r^2F^{\hat{r}}e^{2\Phi})+
{e^{\Phi}\over rsin\theta}{\partial\over{\partial\theta}}(F^{\hat{\theta}}
sin\theta)+
{e^{\Phi}\over
rsin\theta}{\partial\over{\partial\varphi}}F^{\hat{\varphi}}
\end{equation}
It is very easy to see that the above equation again recovers the
well-known spherical symmetric case, as well as the
Newtonian limit.

\section{Physics Inputs and Numerical Algarithms} 

Physics inputs include stellar model, thermal conductivity, 
heat capacity, neutrino emissivity, superfluidity, surface 
temperature treatment, and the position and amount of the 
energy released by a glitch.  In our calculations, we
divide a neutron star into two parts.  The region from the
neutron star center to where the mass density equals the nuclear 
density $\rho_{N}=2.8\times 10^{14}g/cm^3$ is defined as 
core, while the region from $\rho_{N}$ to the boundary density 
$\rho_{b}=10^{9}g/cm^3$ is defined as crust.  The core is 
treated as isothermal with a time-dependent temperature.
In the crust region, 
the temperature after a glitch has both spatial and
temporal variations; hence an 'exact' or 'evolutionary' 
treatment is necessary (Nomoto and Tsuruta 1987, Van Riper 
1991).  In our treatment, the energy flow and the 
spatial dependence 
of the temperature throughout the crust is followed but 
the thermal effects on the stellar structures are neglected, 
i.e. the same hydrostatic stellar model is used throughout the
evolution.

The stellar structure of a neutron star is determinded by the equation of
state (EOS).  We consider three representative EOSs in this paper,
with the total mass of the neutron star in all three models taken to
be $1.4M_{\odot}$ ($M_{\odot}$ is the solar mass).  The first EOS is
BPS (Baym, Pethick, and Sutherland 1971), which is often used as a
soft EOS in neutron star cooling studies.  BPS model results in high
central density and little mass in crust, and hence a small radius.
The second EOS used is PPS (Pandharipande, Pines, and Smith 1976),
which is a stiff EOS with low
central density, thick crust and large radius.  The third EOS UT 
(Wiringa and Fiks 1988), is a representative intermediate
stiff model.  Its central density, crust thickness and radius are
between those of BPS and PPS model.

In the crust region, we fitted the thermal conductivity data 
provided by Van Riper (1991), who followed the work of Itoh 
et al. (1984a,b,c,d), together with the quantum corrections of 
Mitake, Ichimaru, and Itoh (1984).  The ions A and Z which 
required in the conductivity formulae are taken from Lattimer et al. 
(1985, hereafter LLPR).  It is important to note that 
the conductivity decreases as the temperature increases.

For a star with a core temperature $\sim 10^8K$, the important 
neutrino emission processes include electron bremstralung 
(Flowers and Itoh 1976,1979), neutron-neutron, neutron-proton 
bremstralung and the modified Urca process (e.g. Fridman and 
Maxwell 1979).  The rapid cooling processes, e.g. pion 
condensation (e.g. Maxwell 1977) or direct Urca process 
(Prakash et al 1992) etc, will not be considered here.

The heat capacities in the crust region come from 
those of extreme relativistic 
degenerate electrons and those of nonrelativistic neutrons 
and ions.  The capacities of electrons and neutrons 
are given by Glen and sutherland (1980).  
The capacity of ions is given by Van Riper (1991).  
The capacities of the core are mainly due to the 
relativistic electrons, the superconducting protons 
and the superfluid neutrons (Maxwell 1979).  
The transition temperature 
of normal-superfluid neutrons 
is given by Takatsuka and Tamagaki (1971).

We use an empirical formula (Gudmundsson 1983) to relate the
surface temperature $T_{s}$ and the
boundary temperature $T_{b}$ at $\rho_b$
\begin{equation}
\label{eq16}
T_{b8}=1.3(T_{s6}^{4}/g_{s14})^{0.455}\;,
\end{equation}
where $g_{s}$ is the surface gravity.  Eichler and Cheng (1989) 
have shown that the thermal response time between the boundary 
and the surface is of order of seconds, which is even less than our numerical 
time step (greater than 10 s).  The 
above relation is clearly valid in our computation.

In this paper, we study the thermal response of a glitch which
releases heat in a small volume inside the star.  Although the 
stellar structure may be spherical symmetric, the heat transport
is {\it not} spherical symmetric as the heat is deposited off center.
The pulsar glitch may be caused by 
the sudden transfer of angualr momentum from the more rapidly 
rotating crustal superfluid to the solid crust region or the 
sudden fracture of the crustal lattice due to gravitational 
stress or magnetic stress.  The two heating mechanisms have been studied by 
Alpar et al. (1984a,b) and Ruderman (1976,1991).  The 
energy $\Delta E$ is most likely deposited within the density range 
$10^{12}$ to $2.4\times 
10^{14}g/cm^{3}$.  $\Delta E$ is of order $10^{41}\sim 10^{43} ergs$ 
(Van Riper, Epstein, and Miller 1991).  
In this paper, we will focus on the case 
of the Vela pulsar, with $\Delta E=10^{42}ergs$
released at $\sim 10^{12}g/cm^{3}$.

At time $t=0$, the heat is deposited in a small volume at $r=r_g$ and
$\theta=0$.  $r_g$ is the radius where the density is
$\rho_g=10^{12}g/cm^3$ (cf. Fig.1).  Without lose of generality, we
choose the hot spot to be centered at $\theta=0$, which gets rid of
the $\varphi$ dependence in the transport problem.  The transport
equations in the crust region are solved by explicit finite
differencing, while the core is taken to be isothermal with a time
dependent temperature $T_{core}$ determined by the total heat inflow
and the core heat capcity.  In a typical run, the spatial resolution
is taken to be $N_r \times N_{\theta} = 100 \times 50 $, with a
variable time step determined by stability requirements.  Numerical
convergence of the results have been carefully checked.  The
temperature is cell-centered, while the flux is centered on the cell
surface.  The inner and outer boundary conditions for the flux are
given explicitly by
\begin{equation}
\label{eq17}
F^r(1,j)=-Ke^{-\Phi-\Lambda}{e^{\Phi (1,j)}T(1,j)
-e^{\Phi (core)}T_{core}\over\Delta r}~~~~1\le j\le N_{\theta}
\end{equation}
at the inner boundary, and
\begin{equation}
\label{eq18}
F^r(N_r+1,j)=\sigma T_s^4(N_r,j){r_s^2\over r_{ob}^2}
~~~~1\le j\le N_{\theta}  
\end{equation}
at the outer boundary.  $r_{ob}$ and $r_{s}$ are respectively the
radii of the outer boundary of the crust and the star.
$\sigma$ is Stefan-Boltzmann constant, $T_s(N_r,j)$
is the surface temperature of the $j$th angular cell at $r_{ob}$.  The
initial temperature distribution within the crust is that of the
equilibrium state of the same star with the initial core temperature.
The temperature evolution is given by standard finite differencing of 
Eq. (\ref{eq15}).  With a second order scheme no extra boundary condition is
needed for Eq. (\ref{eq15}) as the temperature is cell centered.

\section{Results} 

We take the energy to be released at 
$3\times 10^{11}g/cm^3$
$\le \rho \le 3\times 10^{12}g/cm^3$ with
a solid angle of $2^o\times 2^o$ centered at $\theta=0^o$.  The initial
temperature of the core is taken to be $T_c=10^8K$ and the released
energy $\Delta E=10^{42}ergs$, which is about the energy released by
the Vela pulsar after a giant glitch.  
Figure 1 shows the surface temperature versus the polar angle at
different times.  The hot spot is gradually spreading away from
$\theta=0^o$ and takes about 275 days for the surface temperature to
reach the maximum for a UT star.  The temperature at maximum is about 5 times
higher than the background temperature. However, the hottest region is   
concentrated in a solid angle of $2^o\times 2^o$ centered at $\theta=0^o$.  

Figure 2 shows the
evolution of the surface luminosity as function of time for three
different EOSs, i.e. PPS, UT and BPS.  The peak luminosity is higher, the
time needed for reaching the peak is shorter and the relaxation time is
also shorter for softer EOS in
comparison with the stiff one.  This is because the soft EOS has a much
thinner crust.  We can see that there are substantial differences among 
these EOSs.

Figure 3 compares the luminosity evolutions between
the spherical symmetric case and non-spherical symmetric case.  There
are three major differences between these two cases.  (1) The surface
luminosity of the spherical symmetric case reaches the peak almost 5
times faster than that of the non-spherical symmetric case. 
This results from the fact that the non-spherical symmetric case has
much higher temperature which decreases the conductivity a lot.  (2)
The peak of the total luminosity in the spherical case is lower than
that of the non-spherical symmetric case.  It is because 
in the non-spherical case 
more energy comes out through the surface instead of heating the
core.  (3) Most importantly, although the changes of the total luminosity 
for spherical case and non-spherical case are very small,
the surface temperature of  
the non-spherical symmetric case at a $2^o\times 2^o$ cap area 
changes drastically 
(c.f. Fig.1 and Fig.3c of Chong and Cheng 1994 , hereafter
CC94) while
the surface temperature changes very
little in the spherical symmetric case. In other words, a hot spot should 
show up on the surface of neutron star  200 to 300 days after the glitch
and lasts for about a few hundred days 
if the EOS is soft or intermediate stiff. 

In calculating the evolution of total
luminosity for different core temperatures, we find that the
surface luminosity of the cooler model reaches its maximum
earlier.  This results from the fact that the conductivity is higher
for lower temperature.  The relative increase in the luminosity of the
cooler star is higher than that of the hotter star.  The net increase
of the luminosity of the cooler star is a little higher than that of
the hotter star.  It is because the energy of the glitch spreads to
stellar surface of the cooler star faster than that of the hotter
star.  However, the total energy emitted during the thermal afterglow
period is about the same in these two cases.  
We have also calculated the thermal
evolution inside the neutron star at $\theta=0^o$(cf. Fig.1 of Cheng and Li
1997).  In comparing with
the spherical summetric cases (e.g. Figure 2 of 
CC94), we find that the temperature of the heat pulse is
much higher in this case because the energy is released in a much
smaller volume.  The heat pulse is propagating outward as well as
inward, with a speed slower than that of Figure 2 of CC94 (where the
core temperature is chosen to be $10^6K$, and hence the
conductivity is much larger).

\section{Discussion}
 
Based on a set of general relativistic 
thermal transport and energy balance 
equations, we studied the thermal evolution of a neutron star after
a glitch.  We find that if the energy is released in a 
compact region, a hot spot can appear on the stellar surface.  
For a UT star with an interior temperature $\sim 10^8K$, 
although the surface luminosity only increases by $\sim 10\%$, the 
radiation is emitted from a small area with a temperature 
higher than the background temperature by a factor of $\sim 5$.  
This results in a periodic hard X-ray pulse emission which 
should stand out clearly from the soft 
X-ray background.  A soft EOS greatly enhances this effect and 
a stiff EOS reduces it: the thermal response 
to a glitch can provide important constraints on the EOS.

The time for reaching the peak luminosity is
long, typically $\sim$ 1 year for a UT star, this may make it 
difficult to relate the hard X-ray pulse to the glitch generating it.  
However, since the energy released by each glitch should 
be at a different place of the star, comparing the 
relative pulse phase difference and the intensity variation 
of the hard X-rays observed in different epochs can provide 
evidences for this phenomenon.  Together with a 
detailed spectral analysis, the EOS could be deduced.  
A detailed report on this subject and a comparison with 
observed data will be presented elsewhere.
\clearpage

\begin{center} 
\bf{ Figure Captions} 
\end{center}

Figure 1.--- The thermal profiles of the hot spot at the 
surface of neutron star of UT EOS 
labelled by the time in days after 
a heat input of $\Delta E = 10^{42}ergs$ induced by a glitch 
released at 
$3\times 10^{11}g/cm^3$
$\le \rho \le 3\times 10^{12}g/cm^3$ 
in non-spherical symmetric case 
as a function of polar angle.  
The core temperature is $T_c = 10^8K$.

Figure 2.--- The evolution curves of the total luminosity
for PPS, UT and BPS stars denoted by 
solid, dashed and dot-dashed line respectively.  

Figure 3.--- The evolution curves of the total luminosity of UT 
star in spherical symmetric and non-spherical symmetric cases 
denoted by dashed and solid line respectively.

\clearpage 

\end{document}